\documentclass[twocolumn]{revtex4}
\pdfoutput=1
\usepackage{graphicx}
\usepackage{dcolumn}
\usepackage{longtable}
\usepackage{amsmath}
\usepackage{amssymb}

\begin{document}

\title
{Low energy magnetic radiation enhancement in the f$_{7/2}$ shell}

\author
{S. Karampagia, B. A. Brown and V. Zelevinsky}

\affiliation
{National Superconducting Cyclotron Laboratory and Department of Physics and Astronomy, Michigan State University, East Lansing, MI 48824-1321, USA}

\begin{abstract}

Studies of the $\gamma$-ray strength functions can reveal useful 
information concerning underlying nuclear structure. Accumulated 
experimental data on the strength functions show an enhancement 
in the low $\gamma$ energy region. We have calculated the M1 
strength functions for the $^{49,50}$Cr and $^{48}$V nuclei in 
the $f_{7/2}$ shell-model basis. We find a low-energy enhancement 
for gamma decay similar to that obtained for other nuclei in 
previous studies, but for the first time we are also able to study 
the complete distribution related to M1 emission and absorption. 
We find that M1 strength distribution peaks at zero transition 
energy and falls off exponentially. The height of the peak and the 
slope of the exponential are approximately independent of the nuclei 
studied in this model space and the range of initial angular momenta. 
We show that the slope of the exponential fall off is proportional 
to the energy of the $T=1$ pairing gap.

\end{abstract}

\maketitle

\section{Introduction}

In order to understand the nuclear properties in the quasicontinuum, 
statistical quantities are used, such as the nuclear level density and 
the $\gamma$-ray strength function ($\gamma$SF) \cite{Barth} for a 
particular multipolarity. The strength function is the average reduced 
radiation or absorption probability of photons of given energy $E_{\gamma}$. It is commonly 
adopted that the E1 strength function is dominated by the giant electric 
dipole resonance (GDR) around $E_{\gamma}\approx 78\cdot A^{-1/3}$ MeV, 
which can be reproduced, not too far from the maximum, by a classical 
Lorentz line \cite{Axel,BF}.  It was earlier assumed that the E1 strength 
function for lower energy $\gamma$-rays corresponds to the tail of this 
Lorentzian. Current experimental data \cite{Popov, Voinov} show that the 
Lorentzian description fails for these energies. In order to account for 
the lower $\gamma$ energies, the Kadmenski\u{i}- Markushev-Furman (KMF) 
model \cite{KMF} was suggested. Empirical modifications of this model 
\cite{KMFmod} have also been used to describe the behavior of the E1 
strength function at low $E_{\gamma}$ with the use of the 
temperature-dependent  GDR width.

Experimentally, resonances in the low $E_{\gamma}$ region have long been
observed, commonly termed as pygmy dipole resonances and attributed to the
enhancement of the E1 strength function \cite{pygmy1}, partly due to the 
presence of a neutron skin. Recent studies in rare earth nuclei have
shown \cite{pygmy2, M1enha11} that bumps in the $E_{\gamma} \approx$ 3 MeV  
region are of M1 character. Actually, the M1 transitions seem to play an 
active role in the $\gamma$SF being described also by a Lorentz line 
\cite{M1spinflip} based on the existence of a  resonance that originates 
from spin-flip excitations in the nucleus \cite{BohrII,Heyde}.

In the last decade things have become more complicated, since
measurements of the $\gamma$SF
\cite{M1enha1,M1enha2,M1enha3,M1enha4,Algin,M1enha5,M1enha6,M1enha7,M1enha8,M1enha9,M1enha10,M1enha11,M1enha12,M1enha13}
have revealed a newly observed
minimum around $E_{\gamma} \simeq 2-4$ MeV, so besides the high $E_{\gamma}$
enhancement, there is also a low $E_{\gamma}$ enhancement. The first attempts to
understand the low-$E_{\gamma}$ enhancement \cite{M1enha1,M1enha2,M1enha4} used
the KMF model to describe the GDR; the contribution of the giant magnetic dipole
resonance to the total $\gamma$SF is fitted by a Lorentzian, similarly to
the E2 resonance, while the low-$E_{\gamma}$ region is described by a separate
term that has a power-law parametrization. In \cite{M1enha5} the authors used a functional
form of the $\gamma$SF with contributions from E1 and M1 resonances plus an
exponential low-energy enhancement function to simulate two-step $\gamma$-cascade spectra.
They found that all  M1 strength functions show a low-$E_{\gamma}$
increase compared to the uncertain behavior of the low-energy  E1 strength functions.

In \cite{M1enha9,M1enha12} it was found that the E2 transitions are
of minor importance whereas the dipole transitions dominate in the low-$E_{\gamma}$
enhancement region. The first theoretical evidence  of the strong enhancement at
low $E_{\gamma}$ came from the shell model calculations of $B$(M1) values
for $^{90}$Zr, $^{94-96}$Mo \cite{M1enha8} and $^{56,57}$Fe \cite{Brown} where the 
calculated $B$(M1) and the $\gamma$SF showed large
values for low $E_{\gamma}$. The influence of this low energy enhancement of
the $\gamma$SF is not of minor importance, as it has been found that the neutron capture
reaction rates can grow due to this effect by 1-2 orders of magnitude \cite{Larsen}.

In this study, we calculate $B$(M1)
for $^{49}$Cr, $^{50}$Cr, and $^{48}$V in the model space of $f_{7/2}$ using 
the OXBASH shell model code \cite{Brown3}. Although the model space is small, 
the results lead to new insights. In addition, we are able to consider the M1 
strength for transitions to excited states ($\gamma$ absorption). From this we 
show for the first time that the low-energy part of the M1 distribution is 
peaked at zero energy, and falls off exponentially below and above that point. 
For these nuclei we consider the states with $T=T_z$ obtained with the F742 
Hamiltonian from \cite{Brown2} that reproduces the known low-lying energies
in the nuclei of interest. The results are largely independent of the nucleus, the 
range of initial spins and the excitation energy. We show that the slope of the 
exponential fall off is determined mainly from the $T=1$ (pairing) part of the 
Hamiltonian. 

In the discussion we compare the M1 strength results for the 
$f_{7/2}$ model space with those obtained from the full $pf$ model space for 
$^{48}$V, again for states with $T=T_z$. By allowing the successive occupance 
of all the orbitals of the $pf$ shell, starting with the $f_{7/2}$ orbital 
alone, we explore how the addition of orbitals affects the low-energy 
enhancement and the overall M1 strength disrtibution. We also compare our 
results to the available experimental M1 strength function of $^{50}$V.

\section{Results}

\begin{figure}
\centering
\includegraphics[height=40mm]{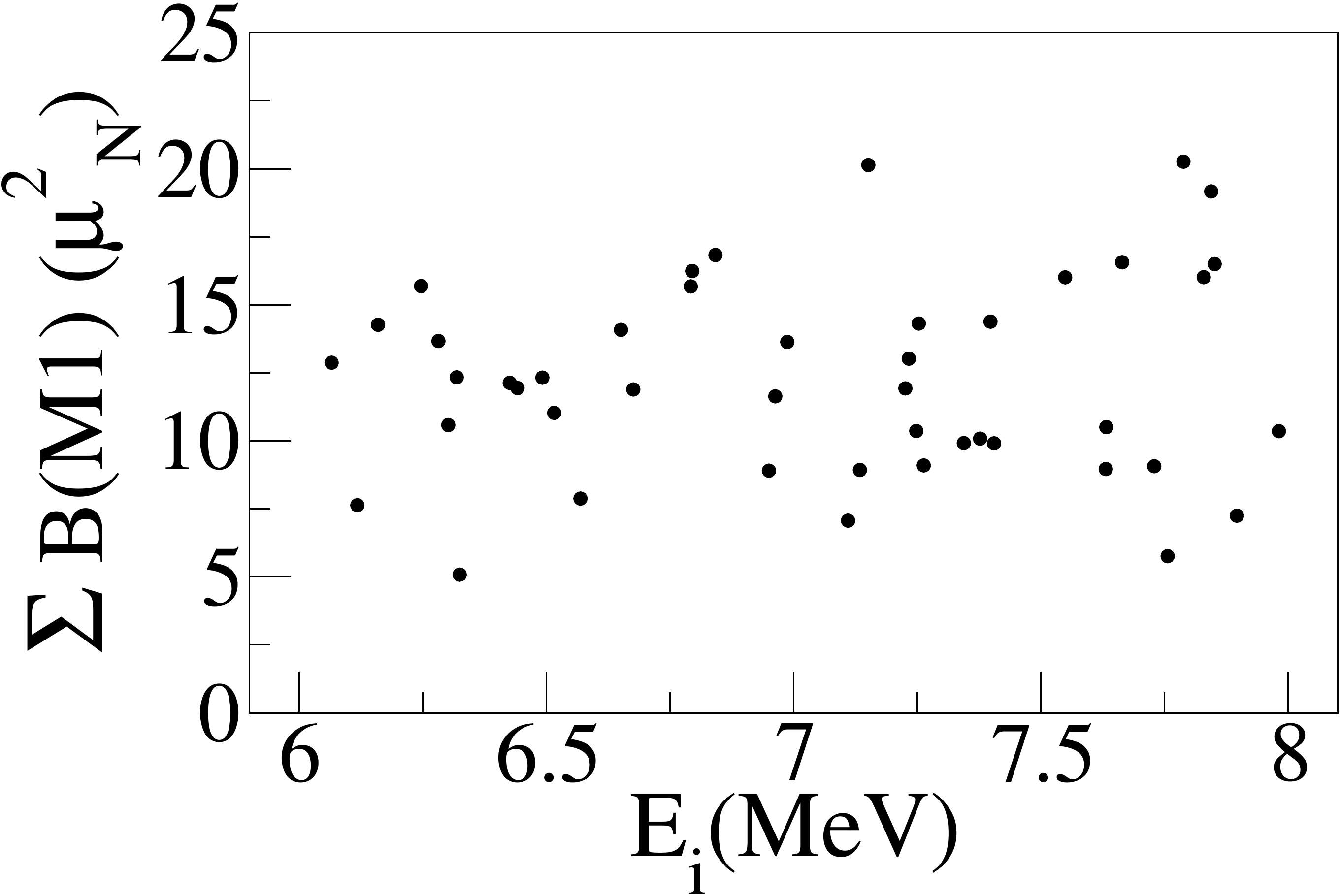}
\caption[]{Summed $B$(M1) strength for a range of initial states in $^{50}$Cr.}    \label{Fig1}
\end{figure}

\begin{figure}
\centering
\includegraphics[height=40mm]{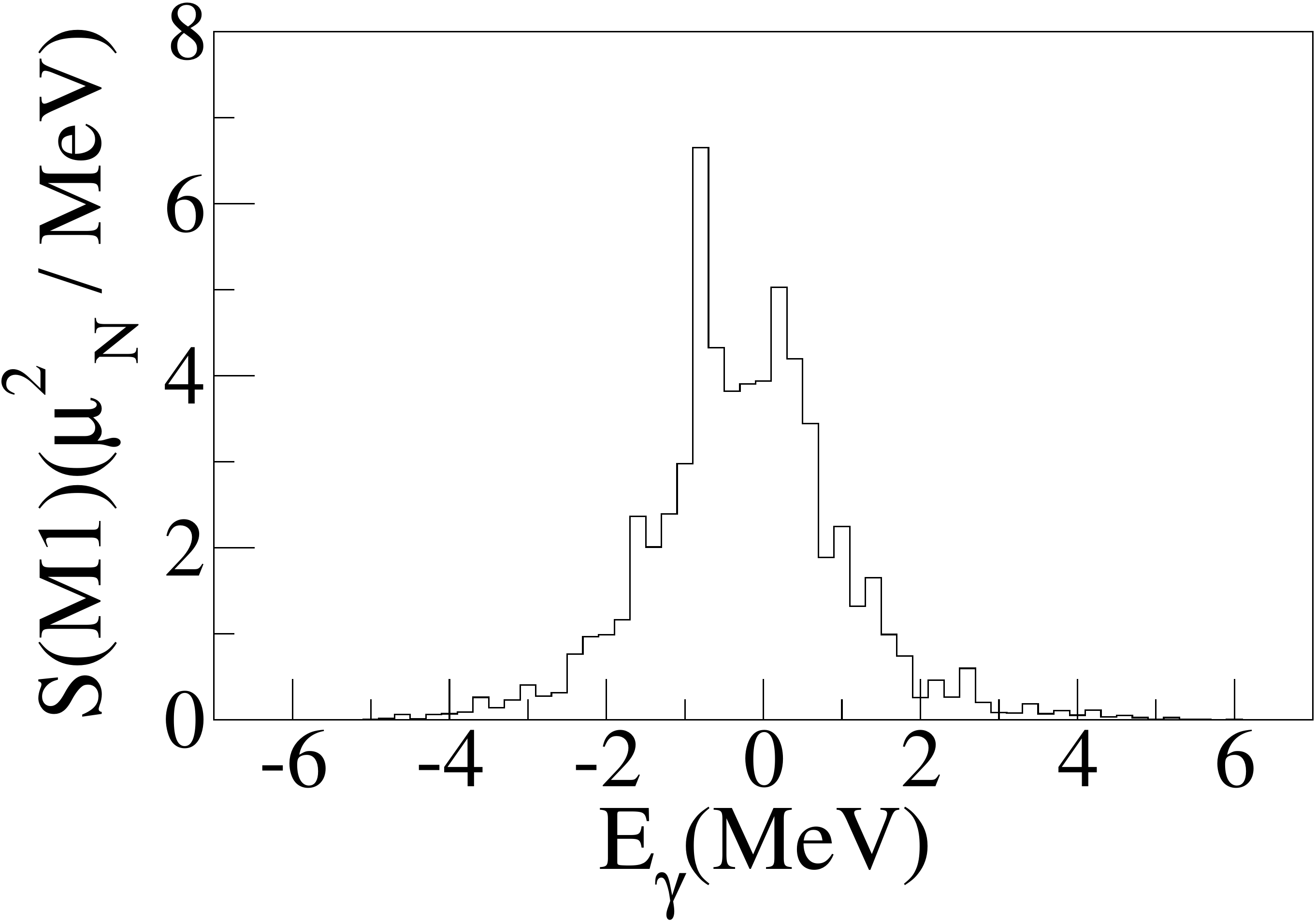}
\caption[]{The average M1 strength distribution for states between 6 and 8 MeV in $^{50}$Cr.}    \label{Fig2}
\end{figure}

We start by considering the states in $^{50}$Cr from 6 to 8 MeV. 
The sum of $B$(M1)s stemming from each initial state is shown in 
Fig. \ref{Fig1}. This has a Porter-Thomas type scatter around an
average value of 12.5 $\mu_N^{2}$. The average M1 strength 
distribution $S$(M1) is shown in Fig. \ref{Fig2}. This is obtained 
by first sorting the $B$(M1)s according to the increasing energy 
differences, $E_{\gamma} = E_i - E_f$ and summing them over bins 
of $\Delta E_{\gamma}$ = 0.2 MeV, for a certain initial energy range (here 
$E_i$ = 6-8 MeV). These are then averaged over the number of initial 
states, 
\begin{equation}
S_i=\frac{\sum_{\text{bins }} B({\rm M1})_{\{ E_i=6-8 \text{MeV} \} }}{\text{Number of initial states}}.
\end{equation}
The area of the $S$(M1) in Fig. \ref{Fig2} is 12.5 $\mu_N^{2}$.
\begin{figure}
\centering
\includegraphics[height=160mm]{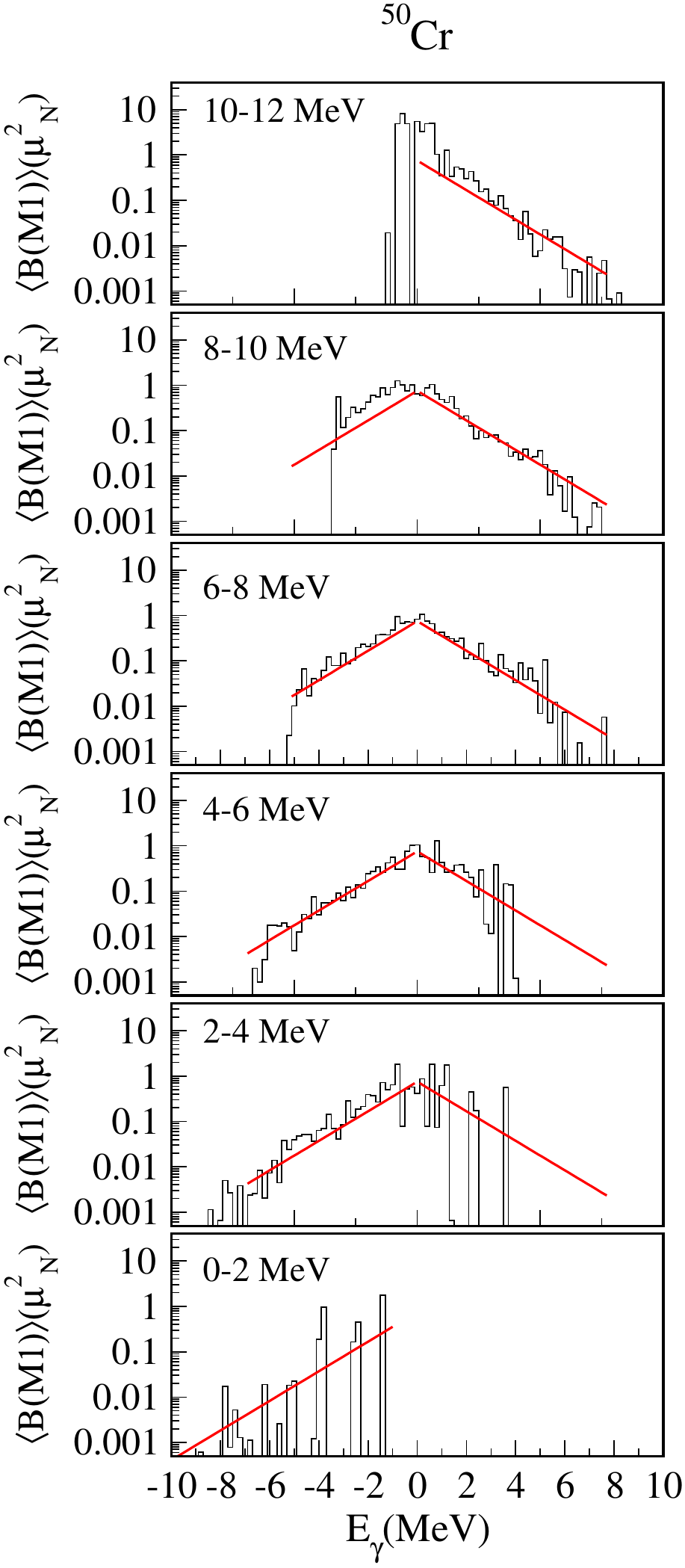}
\caption[]{Average  $B$(M1) values as a function of $\gamma$-ray
energy $E_{\gamma}$ for $^{50}$Cr and initial energies, $E_i$, in various 2 MeV ranges. The lowest
panel is for 0-2 MeV, the highest for 10-12 MeV. Each M1 distribution 
is compared to the same exponential, red line, with parameters $B_0$ = 0.75 $\mu^2_N$ and $T_B=1.33$ MeV.}\label{Fig3}
\end{figure}

\begin{figure}
\centering
\includegraphics[height=100mm]{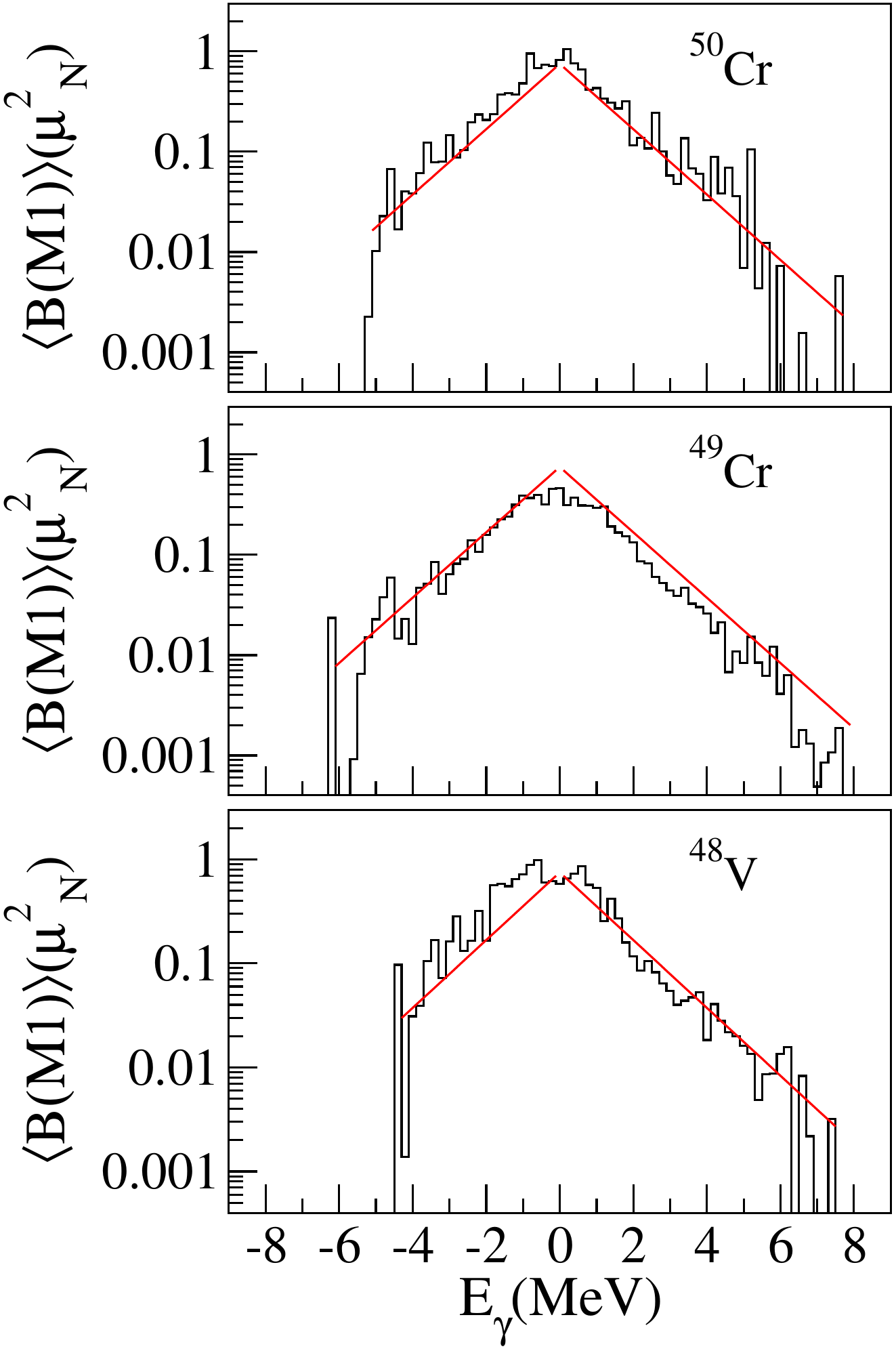}
\caption[]{Average $B$(M1) values as a function of $\gamma$-ray energy $E_{\gamma}$
for $^{49}$Cr, $^{50}$Cr, and $^{48}$V for initial energies, $E_i$, in the interval 6-8 MeV. Each M1 distribution
is compared to the same exponential, red line, with parameters $B_0$ = 0.75 $\mu^2_N$ and $T_B=1.33$ MeV.}  \label{Fig4}
\end{figure}

\begin{figure}
\centering
\includegraphics[height=105mm]{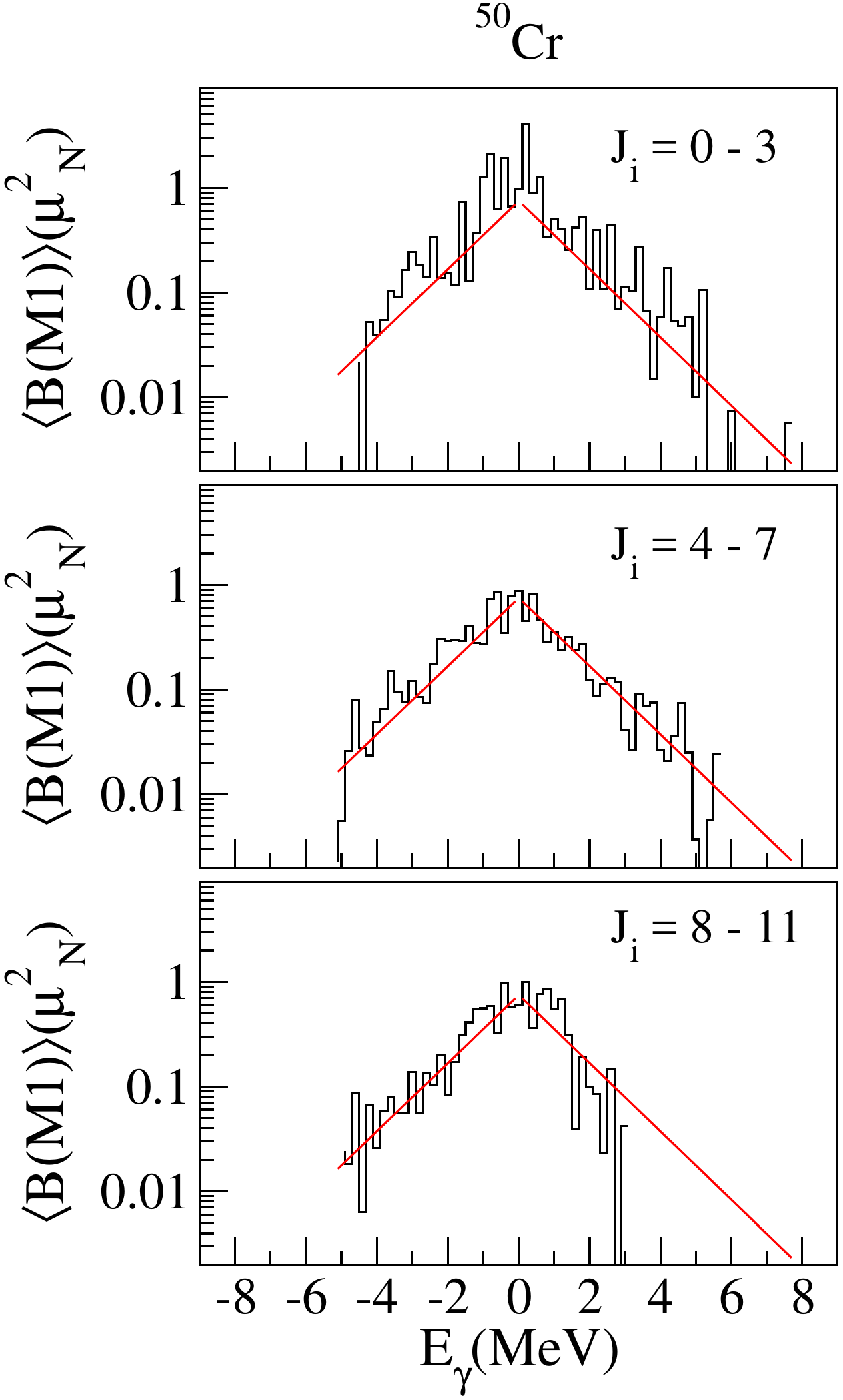}
\caption[]{Average $B$(M1) values for $^{50}$Cr as a function of $\gamma$-ray energy, $E_{\gamma}$, for different initial spin ranges. Each M1 distribution
is compared to the same exponential, red line, with parameters $B_0$ = 0.75 $\mu^2_N$ and $T_B=1.33$ MeV.}\label{Fig5}
\end{figure}

Experimentally, the quantity of interest is the $\gamma$ decay strength
function $\gamma$SF defined by \cite{Barth}
\begin{equation}
f_{{\rm ML}}^i(E_{\gamma})=\rho_{i}\,\frac{\langle\Gamma_{\gamma i}(E_{\gamma})\rangle}
{E_{\gamma}^{2L+1}},                    \label{1}
\end{equation}
where $L$ characterizes the multipolarity of the transition and $\rho_i$  is the
level density of the initial states. The partial radiative width
$\Gamma_{\gamma}$ is given, for M1 transitions, by
\begin{equation}
\Gamma_{\gamma i,M1}(E_{\gamma})=\frac{16\pi}{9}\left(\frac{E_{\gamma}}{\hbar c}\right)^3
B({\rm M1})(E_{\gamma})_i,                         \label{2}
\end{equation}
where the index $i$  specifies selected initial spin values and
the initial energy region $E_i$. By combining the
two expressions we find the $\gamma$SF,
\begin{equation}
f_{{\rm M1}}(E_{\gamma})=a \, \langle B({\rm M1})(E_{\gamma})
\rangle_i\rho_i(E_i),                        \label{3}
\end{equation}
where
\begin{equation}
a = \frac{16\pi}{9(\hbar c)^3}=11.5473 \cdot 10^{-9}\mu_N^{-2} \cdot {\rm MeV}^{-2}. \label{4}
\end{equation}

We will show the results in terms of the $\langle B({\rm M1})(E_{\gamma})\rangle$
of Eq. (\ref{3}). At the end we will consider the $\gamma$SF.
The calculated $B$(M1) values are sorted according to increasing transition
energy, $E_{\gamma}$, and grouped in energy bins of 0.2 MeV width.
For each bin the average $B$(M1) value, $\langle B({\rm M1})(E_{\gamma})\rangle$,
was found by dividing the sum of the $B$(M1) values in this bin by their number.
This leads to a plot whose average value at a given $E_{\gamma}$ does
not depend on the bin size.

The results for $^{50}$Cr are shown in Fig. \ref{Fig3} for several
ranges of initial energies. The straight lines shown in all panels are for the
exponents, $B_0 \, e^{ {-|E_{\gamma}|/T_B} }$, with $B_0=0.75$ $\mu^2_N$ and 
$T_B=1.33$ MeV (the notation of reference \cite{M1enha8} is used). 
A similar exponential behavior is seen in all regions of 
excitation energy, even for the lowest region of 0 to 2 MeV, where only
$\gamma$ absorption can take place. This result is very different from the
Brink-Axel model where the strength function on excited states
is related to the absorption strength function in the ground state.
In contrast, the low-energy distribution is a generic feature for excited
states, that cannot be obtained from information on the ground state
since it peaks at zero energy.

\begin{figure*}
\centering
\includegraphics[height=78mm]{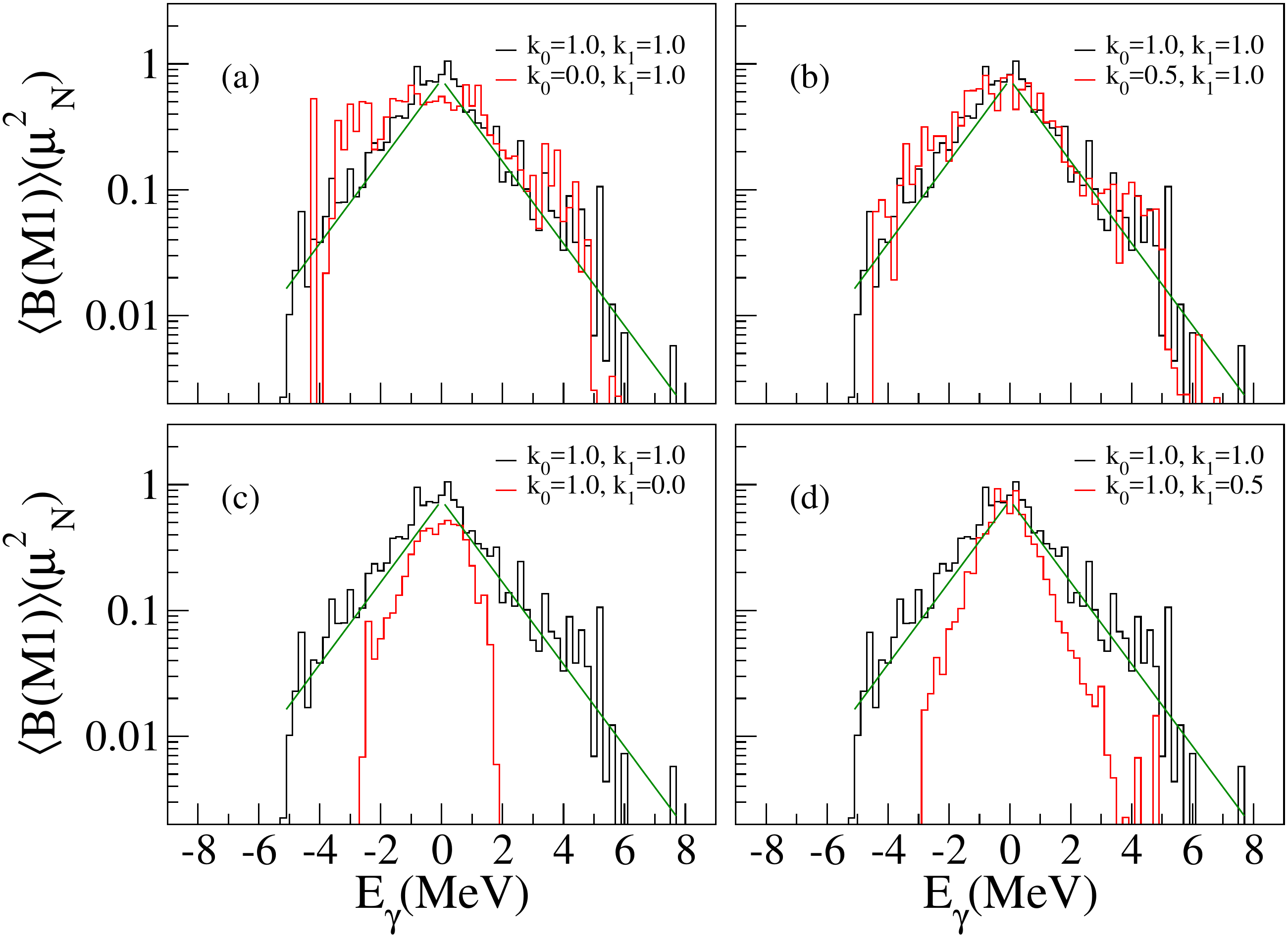}
\caption[]{Average $B$(M1) values as a function of $\gamma$-ray energy $E_{\gamma}$ (black line) for 
$^{50}$Cr for initial energy, $E_i$, in the interval 6-8 MeV, compared with the average $B$(M1) values 
derived using (a) $k_0$=0.0, $k_1$=1.0, (b) $k_0$=0.5, $k_1$=1.0, (c) $k_0$=1.0, $k_1$=0.0, (d)
$k_0$=1.0, $k_1$=0.5, red line. The green line is the exponential fit with 
$B_0$ = 0.75 $\mu^2_N$, $T_B$ = 1.33 MeV.}\label{Fig6}
\end{figure*}

Comparative $B$(M1) diagrams for all three nuclei at $E_i$ = 6-8 MeV can
be seen in Fig. \ref{Fig4}. They all have essentially the same
functional form. The results for $^{50}$Cr divided into different ranges 
for the initial spin are shown in Fig. \ref{Fig5}. The exponential shape 
is independent of spin.

In our orbital space, the two-body interaction Hamiltonian has only eight non-zero matrix elements,
four for the isospin $T=0$ pairs and four for $T=1$. By following the procedure of
\cite{Roman}, we divide the Hamiltonian into two parts and, keeping the symmetry, let
them vary through the numerical coefficients, $k_0$ and $k_1$,
\begin{equation}
H=h+ k_0 V(T=0)+k_1 V(T=1)\label{5},
\end{equation}
where the part $h$ contains the single-particle energies, $V(T=0)$ includes the
matrix elements with $T=0$ while $V(T=1)$ includes the matrix elements with $T=1$.
The absence of the $T=1$ matrix 
elements, (mainly pairing, $J^{\pi}T=$0$^+$1 and $J^{\pi}T=$2$^+$1),
makes the spectrum collapse to low energies. 
We find that the shape of the M1 distribution depends very little on 
the $T=0$ interaction, as shown in Fig. \ref{Fig6}, but there is a strong 
dependence on the strength of the $T=1$ interaction.

\section{Discussion}

For the case of the nuclei studied, it is found that the slope, $T_B$, of the exponential 
functions fitted on the $\langle B({\rm M1})(E_{\gamma})\rangle$, is almost constant 
for all nuclei, while the height seems to vary more, depending on the nucleus. A 
closer look in Fig. \ref{Fig4} shows that the selected $B_0$ value of the preexponent
for $^{49}$Cr slightly overestimates the $\langle B({\rm M1})(E_{\gamma})\rangle$ 
function; however, the choice of a common $B_0$ value for these nuclei gives a good 
description of the $\langle B({\rm M1})(E_{\gamma})\rangle$. 

The approximation of the M1 strength by an exponential function has already been 
proposed in \cite{M1enha8}. There, the $\langle B({\rm M1})(E_{\gamma})\rangle$ 
was calculated using the shell model for $^{94,95,96}$Mo and $^{90}$Zr, in a model 
space which permits both positive and negative parity states. The slope of the 
exponential for the positive parity states ranges from $T_B$=(0.33-0.41) MeV, the lowest 
value corresponding to $^{90}$Zr. The slope of the negative parity states ranges 
from $T_B$=(0.50-0.58) MeV for the Mo isotopes, while $T_B$=0.29 MeV for $^{90}$Zr \cite{M1enha8}. 
The slope for both parities is much more steep than the one found in this study. 

The difference in the exponential slopes in the two studies can be attributed to 
the different orbitals used for the studied nuclei. In our calculations we know that 
it is only the $f_{7/2}$ orbital that contributes to the low$-$energy enhancement, 
but we don't know which are the important orbitals for \cite{M1enha8}. From the text 
it seems that these are the $g_{9/2}$ and $d_{5/2}$, but no further conclusions can 
be drawn. However, we can say that the use of different orbitals will give rise to 
different slopes. Another thing that could be affecting the slope of the low-energy 
enhancement, is the masses of the studied nuclei. As has already been shown, the 
pairing interaction is the main factor that affects the M1 distribution. The 
pairing changes the slope of the $\langle B({\rm M1})(E_{\gamma})\rangle$, in a 
way that, less pairing, gives a steeper slope. Pairing depends on $A$ by a factor 
of $\alpha_p/A^{1/2}$ \cite{Bohr}, so in the $A$=90-96 region, pairing is 25$\%$ 
smaller than the $A$=48-50, thus the slope of the M1 distribution will be steeper. 

\begin{figure}
\centering
\includegraphics[height=55mm]{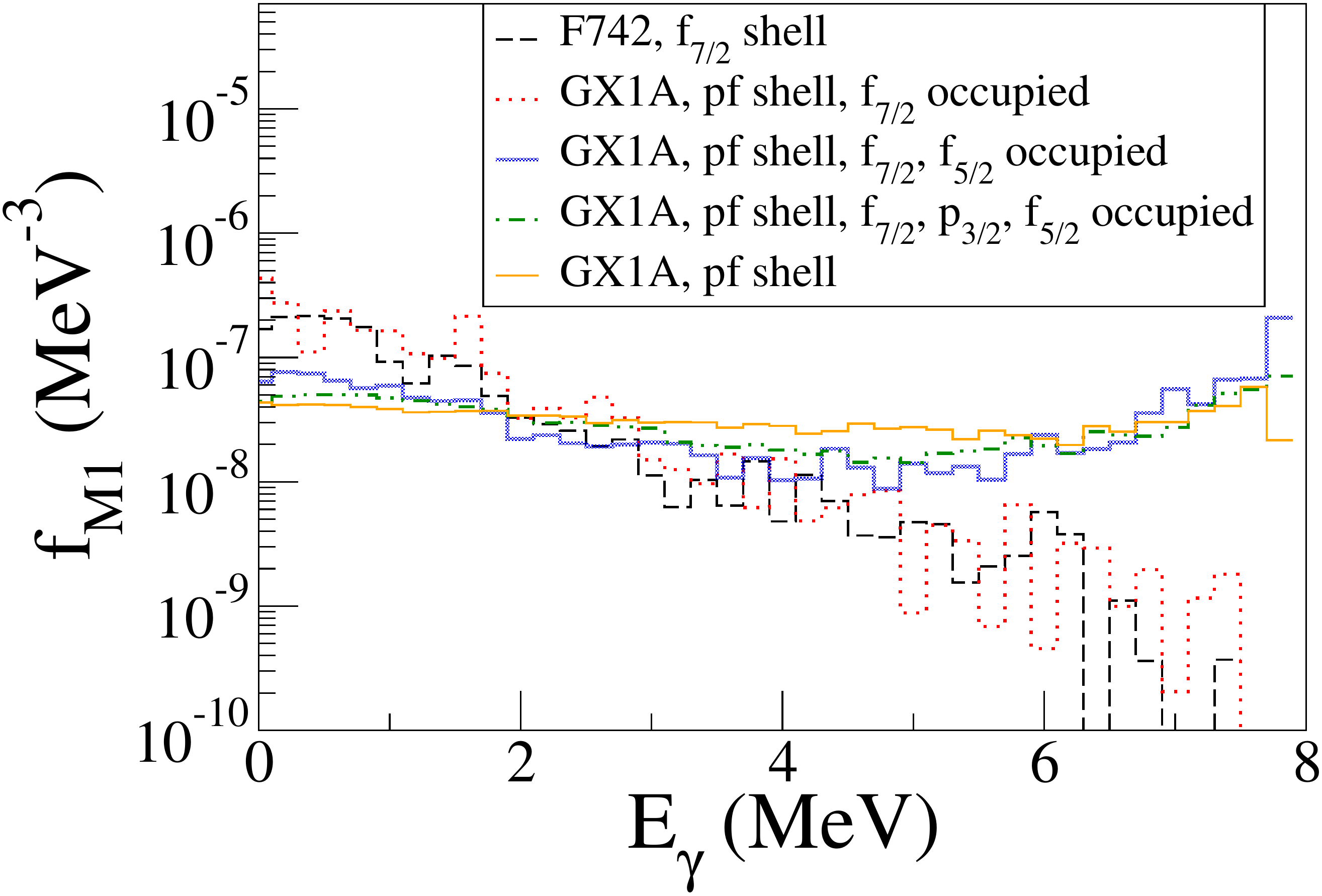}
\caption[]{Calculated $\gamma$SF values for $^{48}$V using the $f_{7/2}$ model space (black 
dashed stair line) and the GX1A interaction in the $pf$ model space allowing first the $f_{7/2}$ 
orbital to be occupied (red dot stair line), then the $f_{7/2}, f_{5/2}$ (blue heavy stair 
line), the  $f_{7/2}, p_{3/2}, f_{5/2}$ (green double dot - slash stair line) orbitals and 
last all the $pf$ shell (orange stair line), as a function of $\gamma$-ray 
energy, $E_{\gamma}$, for initial energies, $E_i$, in the interval 6-8 MeV. }\label{Fig7}
\end{figure}

\begin{figure}
\centering
\includegraphics[height=55mm]{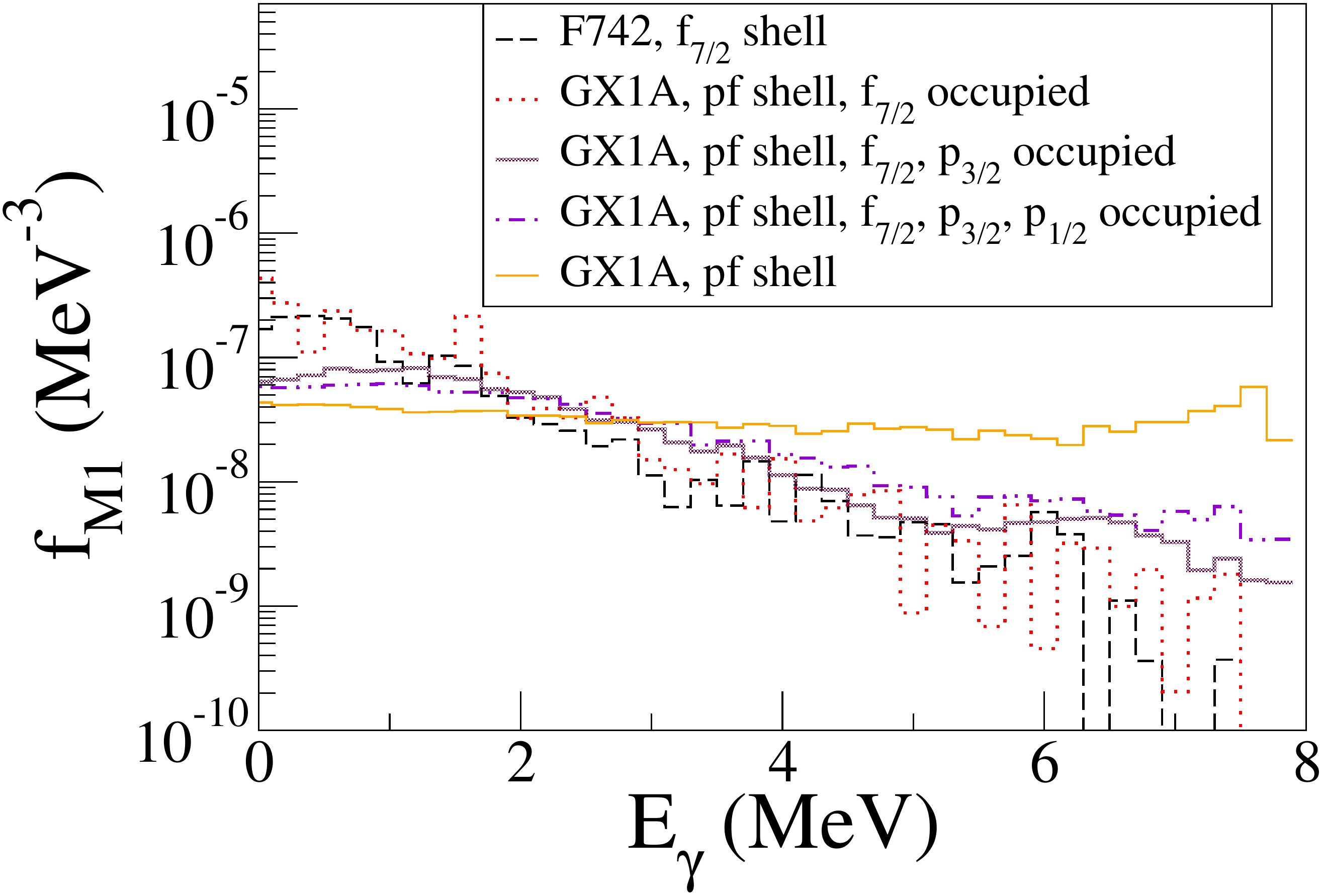}
\caption[]{Calculated $\gamma$SF values for $^{48}$V using the $f_{7/2}$ model space (black 
dashed stair line) and the GX1A interaction in the $pf$ model space allowing first the $f_{7/2}$ 
orbital to be occupied (red dot stair line), then the $f_{7/2}, p_{3/2}$ (violet heavy stair line), 
the  $f_{7/2}, p_{3/2}, p_{1/2}$ (purple double dot - slash stair line) orbitals and 
last all the $pf$ shell (orange stair line), as a function of $\gamma$-ray 
energy, $E_{\gamma}$, for initial energies, $E_i$, in the interval 6-8 MeV. }\label{Fig8}
\end{figure}

In order to explore the point that the consideration of different orbitals will give 
rise to different slopes, we present in Figs. \ref{Fig7}-\ref{Fig8} the calculated 
$\gamma$SF of $^{48}$V from Eq. (\ref{3}), using the $f_{7/2}$ model space (black dashed
stair line) and the GX1A interaction \cite{Honma1,Honma2} in the $pf$ model space, 
allowing successively different orbitals to be added to the model space. In Fig. 
\ref{Fig7} we first allow only the $f_{7/2}$ orbital to be occupied (red dot stair 
line), then the $f_{7/2}, f_{5/2}$ (blue heavy stair line) and $f_{7/2}, p_{3/2}, f_{5/2}$ 
(green double dot - dash stair line) orbitals; finally we compare with the full $pf$ calculation 
(orange stair line). In Fig. \ref{Fig8} we give a different sequence of occupied 
orbitals in the $pf$ model space, starting again with the $f_{7/2}$ orbital (red dot stair 
line), but then allowing the $f_{7/2}, p_{3/2}$ (violet heavy stair line) and $f_{7/2}, 
p_{3/2}, p_{1/2}$ (purple double dot - dash stair line) orbitals to be occupied. We chose to study the 
$\gamma$SF on $^{48}$V because it is the closer nucleus to the available experimental 
$\gamma$SF measurements for $^{50}$V. 

We notice that the full $pf$ shell calculation is more flat compared to the 
$f_{7/2}$ model space or the $pf$ shell calculation, when only the $f_{7/2}$ orbital is 
occupied. In both Figs. \ref{Fig7} and \ref{Fig8}, the successive allowance of occupancy 
of a new orbital makes the $\gamma$SF distribution to drop, up until $E_{\gamma} \sim$ 2 
MeV. For 2 $<E_{\gamma} <$ 4 MeV, the distributions from different occupancies (except 
the full $pf$ calculation) are almost identical. In Fig. \ref{Fig7} we see that the presence 
of the $f_{5/2}$ orbital affects the spectrum for $E_{\gamma} >$ 4 MeV, as it gives a 
spin-flip term which is observed as a peak in the $\gamma$ emission strength, around 
$E_{\gamma}$ = 6-8 MeV. This energy comes from the spin-orbit $f_{7/2}-f_{5/2}$ splitting. 
The addition of more orbitals in the $pf$ model space doesn't change 
the $\gamma$SF for $E_{\gamma} >$ 4 MeV. The effects of the $f_{5/2}$ orbital can be easily 
observed in Fig. \ref{Fig7} as well. There, the successive addition of the $p_{3/2}$ and 
$p_{1/2}$ occupancies doesn't change the $\gamma$SF for $E_{\gamma} >$ 2 MeV. However, the 
addition of the last orbital, $f_{5/2}$, is immediately understood, as the $\gamma$SF 
distribution increases for $E\gamma >$ 4 MeV. The small differences observed for the 
$f_{7/2}$ model space and the $pf$ shell calculation, truncated to the $f_{7/2}$ orbital, 
are attributed to the differences in the interactions, as well as the mass dependence 
present in the GX1A interaction. 

The mixing of the different orbitals with the diagonal $f_{7/2}$ will quench the 
low$-$energy strengths discussed in this study. However, it is mainly the diagonal 
$f_{7/2}$ part which gives the low$-$energy enhancement of the strength function. 
This can also be confirmed by the single-particle occupation numbers of the full 
$pf$ shell calculation. We see that protons and neutrons mainly occupy the $f_{7/2}$ 
single-particle level, the rest of the orbitals having considerably smaller occupation 
numbers. 

\begin{figure}
\centering
\includegraphics[height=60mm]{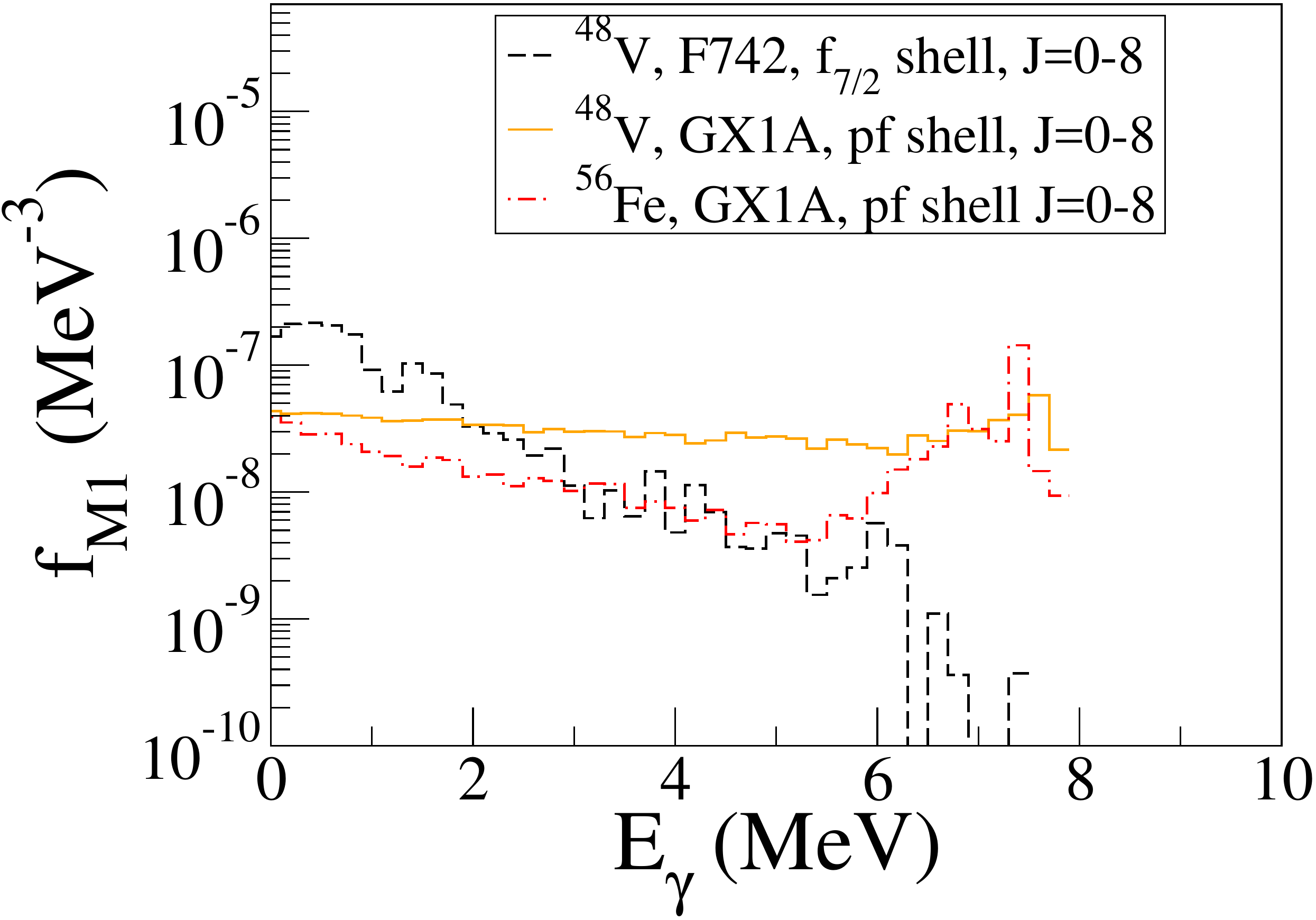}
\caption[]{Calculated $\gamma$SF values for $^{48}$V in the $f_{7/2}$ (black dashed stair line) and the $pf$ (orange stair line) model spaces  compared to the $\gamma$SF values for $^{56}$Fe in the $pf$ (red dot dash stair line) model space, as a function of $\gamma$-ray energy, $E_{\gamma}$, for initial energies, $E_i$, in the interval 6-8 MeV. }\label{Fig9}
\end{figure}

A different example of how the 
mixing of orbitals can affect the M1 strength function can be seen in Fig. \ref{Fig9}. 
There, besides the $^{48}$V calculations using the full $pf$ and $f_{7/2}$ model spaces, 
we also show the M1 strength function of $^{56}$Fe, using a truncated $pf$ space, 
$(0f_{7/2})^{6-t}$$(0f_{7/2},1p_{3/2},1p_{1/2})^{t}$ for protons and 
$(0f_{7/2})^{8-t}$$(0f_{7/2},1p_{3/2},1p_{1/2})^{t+n}$ for neutrons, where $n=2$ and 
$t=$ 0, 1 and 2 \cite{Brown}. We see that the slope of the exponential for 
$E_\gamma \leq 2$ MeV is steeper than the full $pf$ space calculation for $^{48}$V, but  
similar to the $^{48}$V $f_{7/2}$ model space calculation. Further investigation needs 
to be done on how a truncated model space affects the M1 strength function distribution in 
order to fully understand the difference in the slopes of the $pf$ calculations.

\begin{figure}
\centering
\includegraphics[height=60mm]{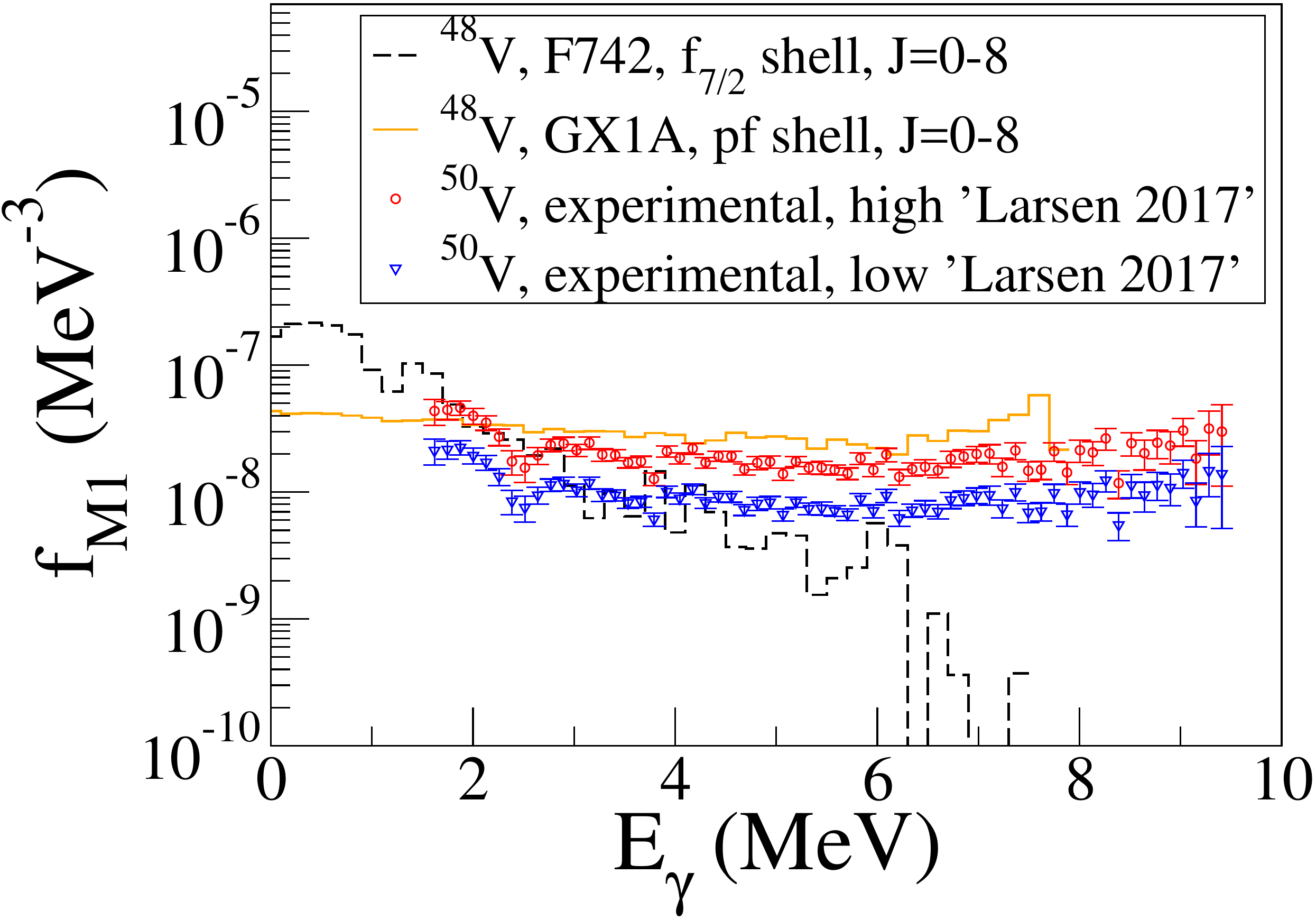}
\caption[]{Calculated $\gamma$SF values for $^{48}$V in the $f_{7/2}$ (black dashed stair line) and the $pf$ (orange stair line) model spaces  compared to the experimental low limit (blue down triangles) and high limit (red circles) data of $^{50}$V (Larsen 2017:
\cite{Cecilie}), as a function of $\gamma$-ray energy, $E_{\gamma}$, for initial energies, $E_i$, in the interval 6-8 MeV. }\label{Fig10}
\end{figure}

The results for the $^{48}$V $\gamma$SF in the $pf$ space (black dashed stair line), along 
with the available experimental data for $^{50}$V (red circles and blue down triangles), are shown in 
Fig. \ref{Fig10}. These data are reanalyzed \cite{Cecilie} and renormalized to new 
neutron-resonance data and new spin distributions. As neutron-resonance data on 
$^{50}$V are not available (since $^{49}$V is unstable), the systematics in this mass region and 
lower/upper limits for $^{51}$V have been used  as constraints. The upper limit of the $^{50}$V 
experimental data agrees better with the theoretical calculations. The lack of experimental 
data below $E_{\gamma}$ = 1.75 MeV makes the comparison with theory difficult in this important 
region. The $\gamma$SFs calculated using the $f_{7/2}$ model space is only added 
for demonstration reasons. As was noted in Figs. \ref{Fig7}-\ref{Fig8}, the $f_{7/2}$ model 
space cannot be used for comparison with the experiment due to the lack of the other orbitals, 
which play also a significant role to the formation of the strength distribution, however it 
can be used to clarify certain physical aspects of the $\gamma$SF.

The exponential form seems to be generic for the problems where we have a bilinear combination of more or less random operators. An analog can be found in the statistical distribution for off-diagonal matrix elements of a realistic many-body Hamiltonian used
in the full shell-model calculations in a finite orbital space. It was studied in detail for an example of the $sd$ shell model long ago \cite{shellPhysRep}, see Figs. 8 and 9 and the Appendix there. Contrary to standard
embedded ensembles of random matrices with Gaussian-like distribution of matrix elements
\cite{Kota}, in such practical applications we typically have a distribution close to the exponential, maybe with some prefactors (mostly important for the smallest matrix elements). This situation
supposedly emerges when the random quantities are matrix elements of multipole operators while the main terms of the many-body Hamiltonian are their bilinear combinations like multipole-multipole forces. Similar to the Porter-Thomas, or more general chi-square, case, the distributions of the bilinear combinations are mainly exponential. The exponential factor, as the effective temperature above, can be roughly estimated as the mean (over the spectrum) excitation energy characteristic for the multipole operator. In our small orbital space, the spin-orbital and monopole terms are reduced to constants. The effective Hamiltonian governed by the pairing-type interaction contains also less coherent parts creating actual superpositions corresponding to complicated stationary states. The diagonal in seniority matrix elements of a time-odd operator, such as the magnetic moment, are not renormalized by pairing. This corresponds to the maximum strength at small $E_{\gamma}$. For the components changing the seniority the mean transition energy is of the order of the pairing gap $\Delta$ equal to about 1.5 MeV for this group of nuclei. This  estimate agrees with the effective temperature $T_B$ found above.

This physics cannot satisfy the Brink-Axel hypothesis which can be approximately valid for the excitations of general
macroscopic nature. In the GDR case, the main part is played by the local dipole polarization of the nuclear
medium which is essentially a universal property of nuclear matter. Such an excitation can be erected on top 
of any shell-model state. In the case considered above, low-energy properties, such as isovector pairing and spin-orbit 
splitting of specific single-particle orbitals, are crucial.

\section{Conclusion}

Summarizing, we have performed shell-model calculations in the $f_{7/2}$ shell,
producing the full spectra and decay schemes of $^{48}$V, $^{49}$Cr, and $^{50}$Cr.
The results indicate a strong low-$E_{\gamma}$ $B$(M1) component, in accordance 
with experimental and theoretical findings. The new outcome of this study is that 
the low energy enhancement is essentially a one-partition phenomenon. Also, it is 
practically independent of the initial energy window or the spin distribution 
considered. All the $B$(M1) functions can be well fitted as exponential, while it 
is shown that it is the $T=1$ matrix elements which are responsible for the 
exponential shape (the $T=0$ matrix elements provide a very small bump at low 
energies). The comparison of the calculations of the $\gamma$SF in the $f_{7/2}$
and the full $pf$ shell model space, as well as for the successive occupation of different 
orbitals in the $pf$ model space, suggests that the mixing of different orbitals 
with the $f_{7/2}$ leads to the quenching of the low$-$energy enhancement. The 
$f_{5/2}$ orbital has a special role, as it gives a spin$-$flip peak at 
$E_{\gamma}$ = 6$-$8 MeV.  The role of spin-orbital interactions should be studied
in more detail.  \\
\\

\section*{Acknowledgements}

We acknowledge support from NSF grant PHY-1404442.

\end{document}